# Resolving the "Thermal Labyrinth" inside 3D-IC Stacks

Xufei Guo[1,2,#], Yiyun He[1,#], Zirou Chen[1], Zifeng Huang[1], Runsheng Wang[1], and Zhe Cheng[1,*]
[1]School of Integrated Circuits, [2]School of Electronics Engineering and Computer Science, Peking University, Beijing, China
[*]Email: zhe.cheng@pku.edu.cn [#]These authors contributed equally to this work.

***Abstract*—**The heat pathways in 3D-ICs have long been a "labyrinth" lacking experimental investigations into spatial distribution of thermal properties, severely limiting study of internal thermal management inside chips. For the first time, we established a methodology for 3D thermal characterizations of advanced 3D-ICs, implemented through a systematic workflow that enables layer-resolved thermal property extraction and the construction of a location-specific thermal database. We demonstrated its application to an advanced sample featuring 2.2-µm-diameter TSVs and ~2-µm-pitch face-to-face hybrid bonding (HB). Time-domain and steady-state thermoreflectance were used to extract the thermal conductivities (TCs) of Al traces, TSV, BEOL interconnects and HB pads, indicating HB structure enables efficient thermal transport with >9 W/m-K TC. Further, we mapped and visualized the TC distribution of TSV structures and RDL Cu wires with small step sizes down to 0.2 µm. Utilizing these results, we proposed an experiment-based thermal model to reveal that co-evolution of interconnect and substrate TCs is essential for thermal design and fabrication towards effective cooling optimization.

## I. Introduction

3D integration has been a definite trend for the industry as it benefits from higher interconnect density and lower delay (**Fig. 1**). Since interconnect constitutes the core of 3D-IC technology, conventional cooling schemes relying on external heat removal outside of macro-scale packages are no longer sufficient to tackle pressing thermal issues in high-power applications like AI and high-performance computing [1], [2]. Internal thermal management of heat conduction inside 3D-ICs has become as critical [3], [4], [5] but still lacks experimentally determined thermal properties especially for novel interconnect structures (i.e., TSVs and bonding pads) and dielectric layers (i.e., bonding dielectric). Reliable thermal simulation and subsequent cooling optimization would be impossible if such a "labyrinth" of heat pathways remains unresolved, as thermal conductivities (TCs) of nano- and micro-scale structures are often heavily degraded compared to bulk materials due to crystal boundaries and heterogeneous interfaces. Previous research has found a large discrepancy between the actual low TC of sub-µm BEOL layers and their high predicted values provided thermal boundary conductance (TBC) is not taken into account [3].

To address this challenge, we proposed an experiment-based methodology for **3D thermal characterizations of advanced 3D-ICs** (**Fig. 2**), which bridges local thermal measurements and system-level thermal analysis by experimentally reconstructing the thermal properties of the **complete heat conduction pathway**. Layer-resolved thermal properties obtained from representative material layers, interconnect structures, TSV and hybrid bonding (HB) pads were systematically integrated into a **unified 3D-IC thermal database** and subsequently translated into experimentally calibrated finite element thermal models. Beyond the specific stack investigated in this work, the proposed methodology provides a general workflow applicable to a broader range of advanced 3D integrated technologies. By replacing assumed material parameters with experimentally established thermal properties, it enables quantitative hotspot analysis, cooling optimization, and thermal-aware 3D-IC design based on physically grounded thermal models.

## II. Sample Preparation and Thermal Characterizations

We present here the first experiment-based 3D TC distribution in a TSV-embedded, hybrid-bonded 3D-IC stack as the first demonstration of the proposed methodology, leveraging time-domain thermoreflectance (TDTR) and steady-state thermoreflectance (SSTR) techniques. **High-resolution mapping** was also employed to reveal the spatial TC variation in particular regions like TSVs and RDL Cu wires.

The sample in this work consists of two dies bonded face-to-face, designated as Die 1 and Die 2. TSVs were placed in the backside of Die 1 connecting to Al traces at the chip surface and Cu RDL layers near the HB interface (**Fig. 1**). The sample was delayered and precision polished to expose different layers of structures (**Fig. 3**). Besides, a shallow-angle grinding and polishing technique was employed to expose the entire HB interface across the chip area (**Fig. 3(e)**). Regarding thermal characterization methods, TDTR is a pump-probe technique capable of measuring TC at µm-scale resolution and TBC in layered structures [6], [7] (**Fig. 4(b)**). SSTR is a steady-state technique that probes deeper into the sample to extract effective TC over a large volume (**Fig. 4(a)**).

## III. Results and Discussion

By preparing and examining over 10 samples exposing different depths of 3D-IC structures, we recognized evident size-dependent TC evolution of interconnect structures and identified interlayer dielectric (ILD) layers and thin BEOL interconnects as primary thermal bottlenecks with the lowest TCs on the heat pathway.

Surface Al traces and corresponding SiN/SiO ILD were first exposed following the removal of the passivation layer (**Fig. 5**). TDTR single-point measurements found that the TC of Al (averaging 173 W/m-K) is lower than that of bulk Al (237 W/m-K). On the other hand, the TC of SiN and SiO dielectrics (1-2 W/m-K) are consistent with literature values [8]. Following the removal of Al traces, **Fig. 6** shows the cross-sectional schematic (**Fig. 6(a)**) and plan-view images (**Fig. 6(b-c)**) of TSVs buried in top dielectric and Si substrate. TDTR measurements revealed a uniform TC distribution in Si averaging 127 W/m-K (**Fig. 7**), slightly lower than that of undoped bulk Si (142 W/m-K) [9]. The TC of surrounding SiO dielectric is found to be ~1.2 W/m-K. Moreover, the TC of TSV was measured with a 20× objective lens to accurately overlap a smaller beam spot with the centers of TSVs. The measured TC values average 72 W/m-K, lower than previous reports on electroplated Cu (127 W/m-K) [10]. It is worth noting that mapped results in **Fig. 6(e)** achieved excellent consistency with TSV geometry by leveraging a fine scanning step of 0.2 μm and a spatial resolution of ~1.7 μm based on laser spot sizes. The stacked thermal resistance of the region above the interconnect layers in Die 1, namely the majority of the volume containing TSVs, was also measured by SSTR to be 2.4 $m^2$-K/MW, which is consistent with TDTR measurements.

Further, the BEOL interconnect stack in Die 1 were exposed primarily by plasma focused ion beam (PFIB). SEM and TEM images (**Figs. 8-10**) were used to identify metal/dielectric distribution within the remaining layers, which was in turn used for calculating the weighted effective volumetric heat capacity $C_{V,eff}$ (**Fig. 9**). The TCs of these interconnect layers (several tens to hundreds of nanometers thick) are very low (1-3 W/m-K), which agrees with literature [3], [4] and could be attributed to the great number of metal-dielectric interfaces. As measured in Die 2 (**Figs. 11-12**), thicker RDL Cu wires (~1 μm thick) have TCs between 70-90 W/m-K, while dummy regions in between wires exhibit significantly lower TC (3-6 W/m-K).

Another key source of thermal resistance, namely the HB region consisting of dielectric and electrically connected (functional) or isolated (dummy) Cu pads, is exposed following the removal of Die 1 (**Fig. 13**). TDTR measurements reveal relatively high TCs of the bonding layer (>9 W/m-K) in both connected and unconnected regions, which could be attributed to the high density of μm-scale Cu pads and high TBCs of both Cu-Cu and dielectric-dielectric interfaces. While previous studies on HB thermal characterizations mainly rely on test vehicles with larger pad sizes [11], [12], this work enables TC measurement of dense HB areas directly in 3D-IC samples.

## IV. Thermal Simulation and Optimization

Exploiting measured 3D TC distribution in our 3D-IC stack, we were able to establish an **experiment-based thermal model** and perform finite element simulations to evaluate its thermal behavior (**Fig. 14**). The periodic unit model includes two stacked dies, interconnect layers, and the HB layer. On top of setting uniform power maps in Dies 1 and 2 (100×100 $\mu m^2$, 50 and 10 $W/cm^2$ respectively), we also added a localized hotspot (10×10 $\mu m^2$, 1,000 $W/cm^2$) to Die 1 for simulation of non-uniform heat dissipation. Si, SiC, and diamond were used as substrate materials to investigate the effect of substrate thermal conductivity.

**Fig. 15(a)** shows that the peak temperature rise increases with hotspot power density. Replacing Si with SiC or diamond reduces the peak temperature rise, with diamond showing the best thermal performance. However, the smaller improvement from SiC to diamond indicates that the heat-spreading benefit gradually saturates, and the thermal bottleneck also exists in the interconnect and bonding layers. Therefore, a scaling factor $\eta$ was applied to the thermal conductivities of the interconnect and HB layers (**Fig. 15(b)**). Increasing $\eta$ reduces the peak temperature rise, confirming the importance of near-junction layers in heat dissipation. Such improvement can be achieved by using higher-TC dielectrics, increasing metal density, or optimizing the interconnect layout. The reduced benefit at large $\eta$ indicates that other stacked layers and the cooling boundary become the dominant thermal limitations. These results indicate that co-evolution of substrates and near-junction layers towards higher TC is critical for 3D-IC cooling optimization.

From a methodological point of view (**Table I**), this work combines characterization, simulation and optimization so as to extend previous experiment-based thermal characterization studies from individual structures to complete heat conduction pathways in advanced 3D-ICs. We stress that repeatable sample preparation techniques and unified thermal fitting models are necessary to obtain meaningful and comparable results.

## V. Conclusion

This work provides the first demonstration of 3D thermal characterizations and cooling optimization within a TSV-embedded and hybrid-bonded 3D-IC. The HB regions exhibit superior structural quality, enabling efficient thermal transport (>9 W/m-K). Overall, we established a practical methodology and workflow to resolve the "thermal labyrinth" in complex 3D architectures by reconstructing complete heat pathways.

## Acknowledgment

The authors acknowledge the financial support from the National Key Research and Development Program of China (Grant No. 2024YFA1207901) and the National Natural Science Foundation of China (NSFC) (Grant Nos. 62574007, T2550270).

## References

[1] M. S. Lundstrom *et al.*, *Science*, vol. 378, no. 6621, pp. 722–723, 2022, doi: 10.1126/science.ade2191. [2] K. Kim *et al.*, *ACS Nano*, vol. 20, no. 24, pp. 17119-17142, 2026, doi: 10.1021/acsnano.6c01518. [3] Z. Huang *et al.*, *IEDM*, 2025, doi: 10.1109/IEDM50572.2025.11353753. [4] Y. He *et al.*, *EDTM*, 2026, doi: 10.1109/EDTM65772.2026.11497295. [5] Y. Shen *et al.*, *International Communications in Heat and Mass Transfer*, vol. 178, p. 111751, 2026, doi: 10.1016/j.icheatmasstransfer.2026.111751. [6] D. G. Cahill, *Rev. Sci. Instrum.*, vol. 75, no. 12, pp. 5119–5122, 2004, doi: 10.1063/1.1819431. [7] Z. Cheng *et al.*, *Appl. Phys. Rev.*, vol. 11, no. 4, p. 041324, 2024, doi: 10.1063/5.0185305. [8] D. G. Cahill, *Rev. Sci. Instrum.*, vol. 61, no. 2, pp. 802–808, 1990, doi: 10.1063/1.1141498. [9] H. R. Shanks *et al.*, *Phys. Rev.*, vol. 130, no. 5, pp. 1743–1748, 1963, doi: 10.1103/PhysRev.130.1743. [10] T. Gebrael *et al.*, *Nat. Electron.*, vol. 5, no. 6, pp. 394–402, 2022, doi: 10.1038/s41928-022-00748-4. [11] S.-H. Lee *et al.*, *Electronics*, vol. 14, no. 13, p. 2682, 2025, doi: 10.3390/electronics14132682. [12] B. Nili *et al.*, *ECTC*, 2026, pp. 1627–1632. doi: 10.1109/ECTC51846.2026.00266.

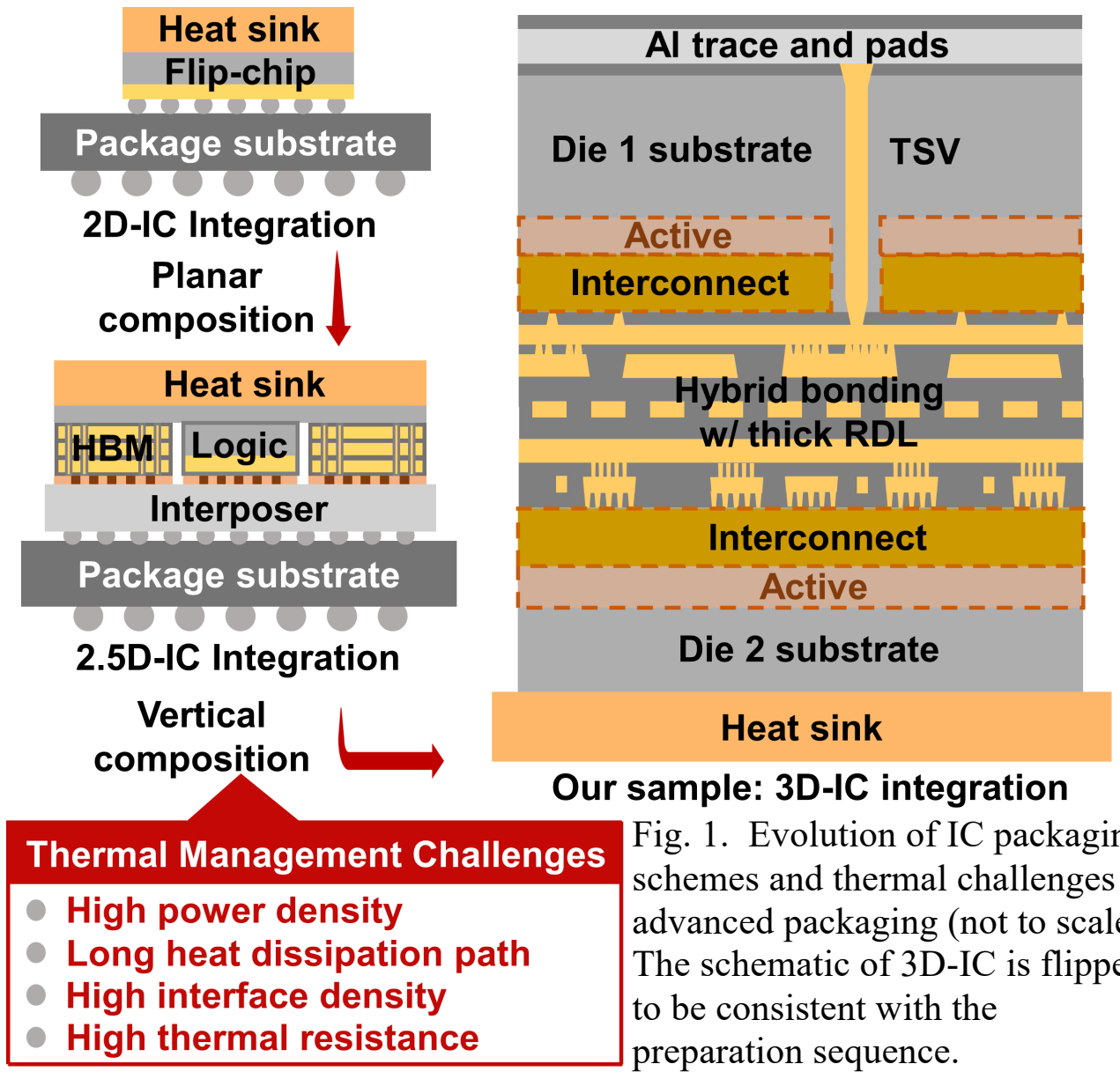

Fig. 1. Evolution of IC packaging schemes and thermal challenges in advanced packaging (not to scale). The schematic of 3D-IC is flipped to be consistent with the preparation sequence.

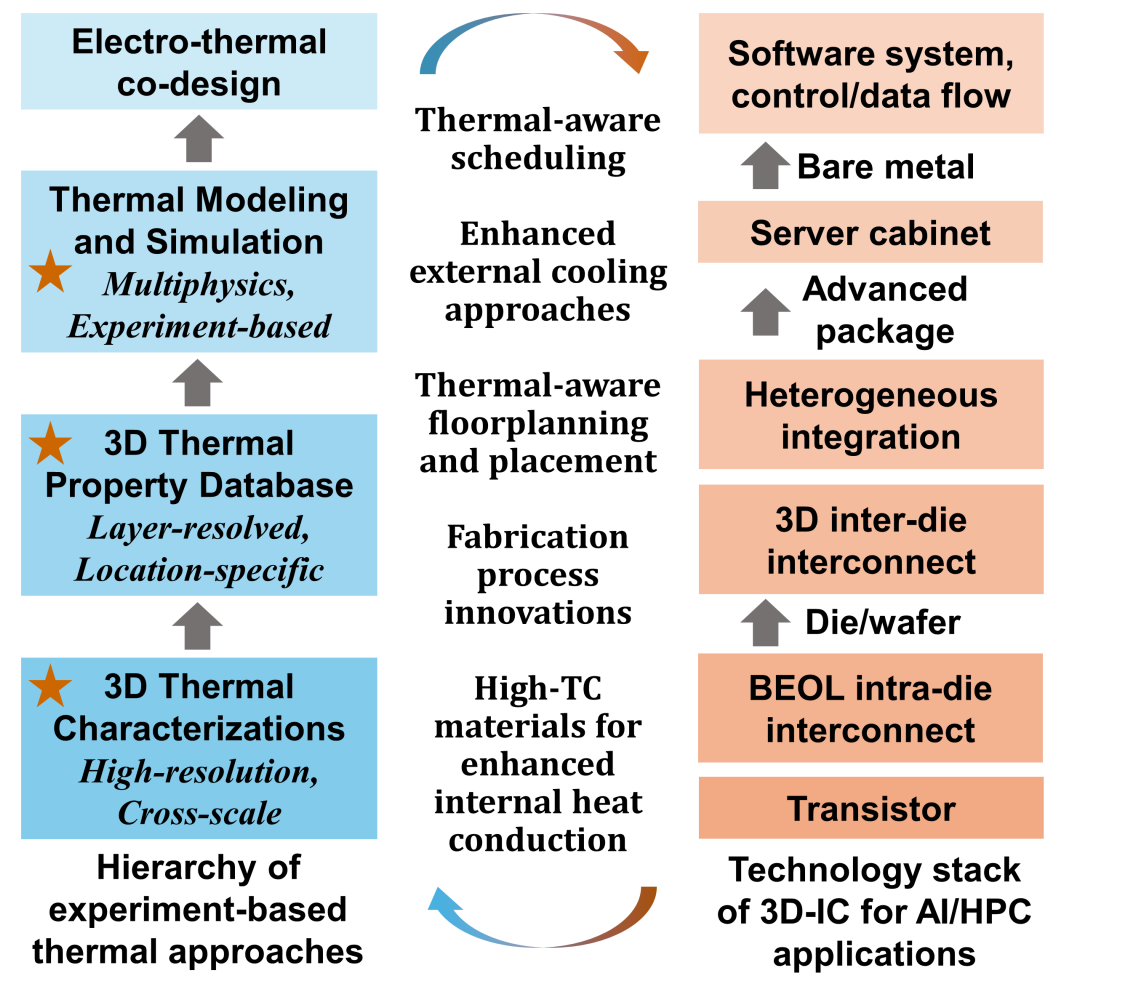

Fig. 2. General workflow of the proposed methodology in this work, aimed at utilizing a hierarchy of experiment-based thermal approaches to solve the pressing thermal issues in different levels of the 3D-IC technology stack. Star marks indicate the scope of examples in this research, the result of which should become the numerical foundation of subsequent electro-thermal co-design paradigms.

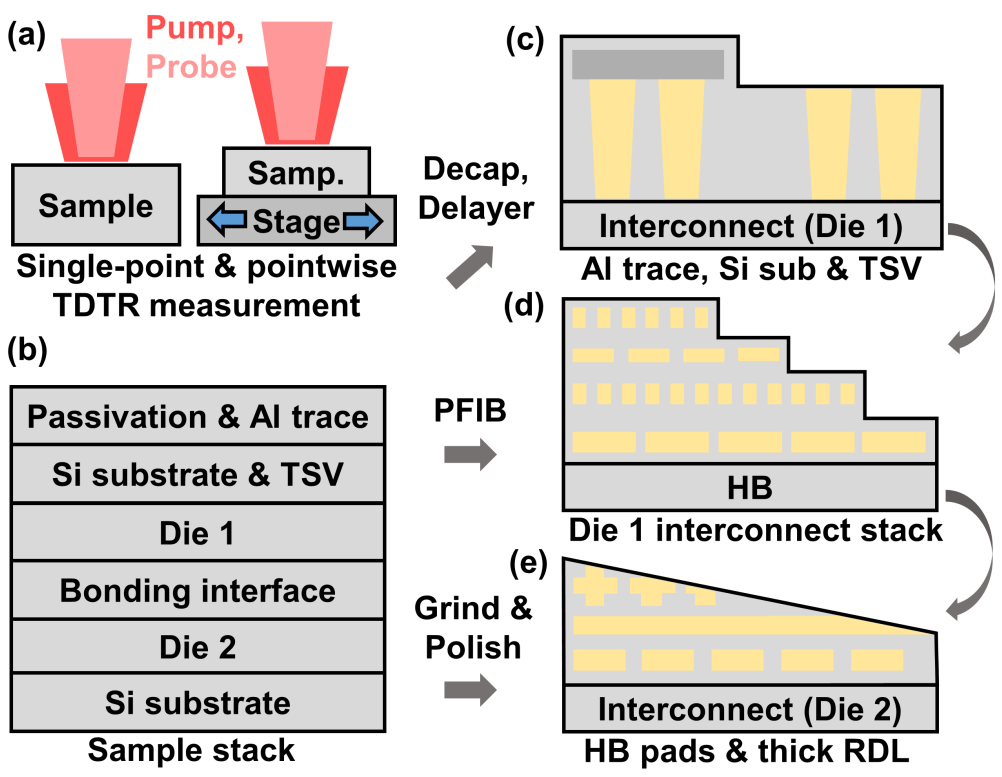

Fig. 3. (a) Thermal characterization schemes and (b) sample schematic (not to scale) with (c-e) exposed surfaces through different sample preparation techniques. The surfaces were protected immediately after being exposed. They were later coated with Al thin films as the transducer layer for TDTR/SSTR measurements.

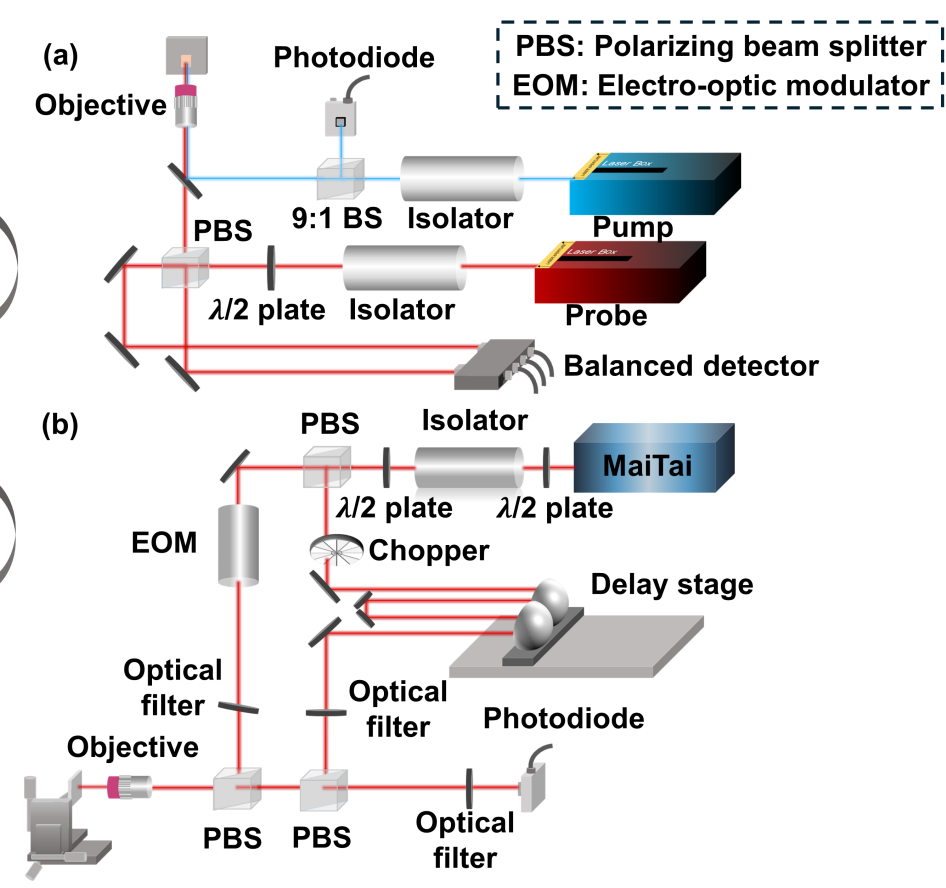

Fig. 4. Simplified setup diagrams of thermal characterization platforms. (a) The SSTR system. (b) The two-tint TDTR system capable of high-resolution mapping measurements.

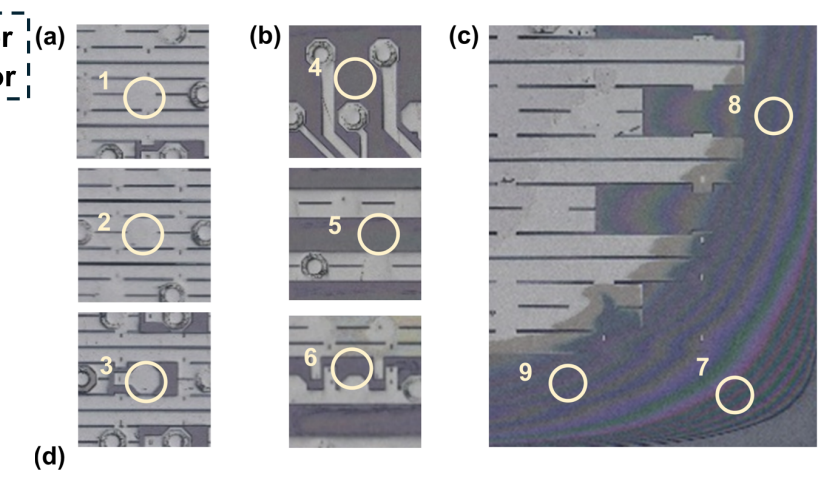

(d)

| Point (Al) | TC (W/m-K) | Point (SiN) | TC (W/m-K) | Point (SiO) | TC (W/m-K) |
|---|---|---|---|---|---|
| #1 | 179 | #4 | 1.8 | #7 | 1.4 |
| #2 | 180 | #5 | 1.7 | #8 | 1.3 |
| #3 | 160 | #6 | 1.8 | #9 | 1.2 |

Fig. 5. (a-c) Optical microscopy (OM) images (before transducer deposition) of (a) surface Al traces, (b) SiN dielectric and (c) SiO dielectric. (d) TDTR-measured TC values of different regions including both metal and ILD.

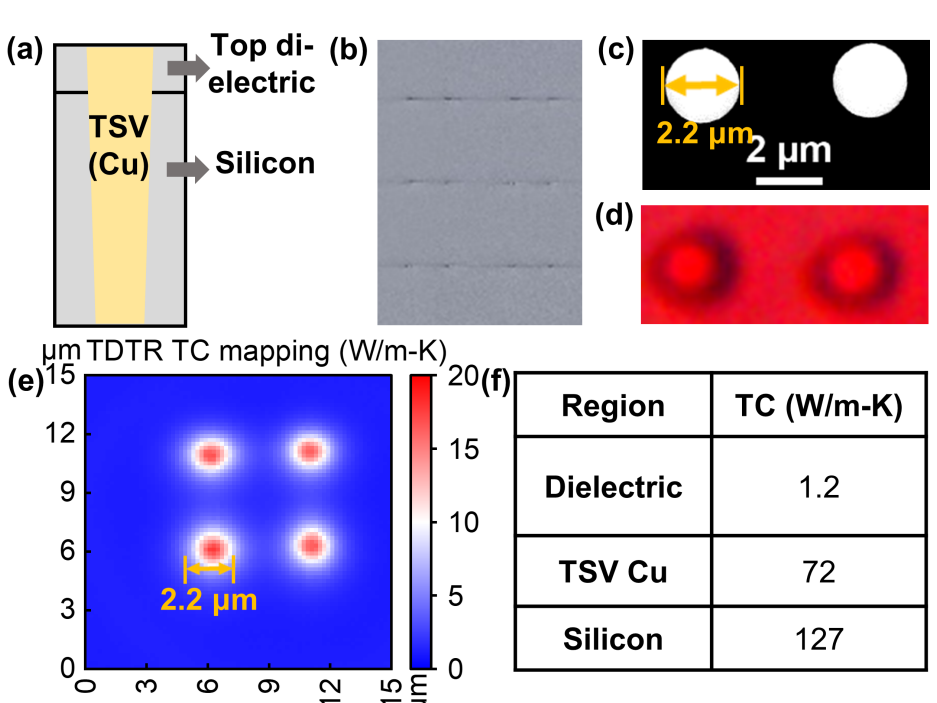

(f)

| Region | TC (W/m-K) |
|---|---|
| Dielectric | 1.2 |
| TSV Cu | 72 |
| Silicon | 127 |

Fig. 6. (a) Regions of interest in the TSV structure, along with (b) OM, (c) SEM and (d) CCD plan-view images. (e) Visualization of spot-averaged TDTR TC distribution of TSVs surrounded by dielectric. (f) Fitted TCs of dielectric, Cu and Si from single-point TDTR measurements. The local TC is a weighted average of the TCs of Cu and the dielectric, from which the TC of Cu is determined.

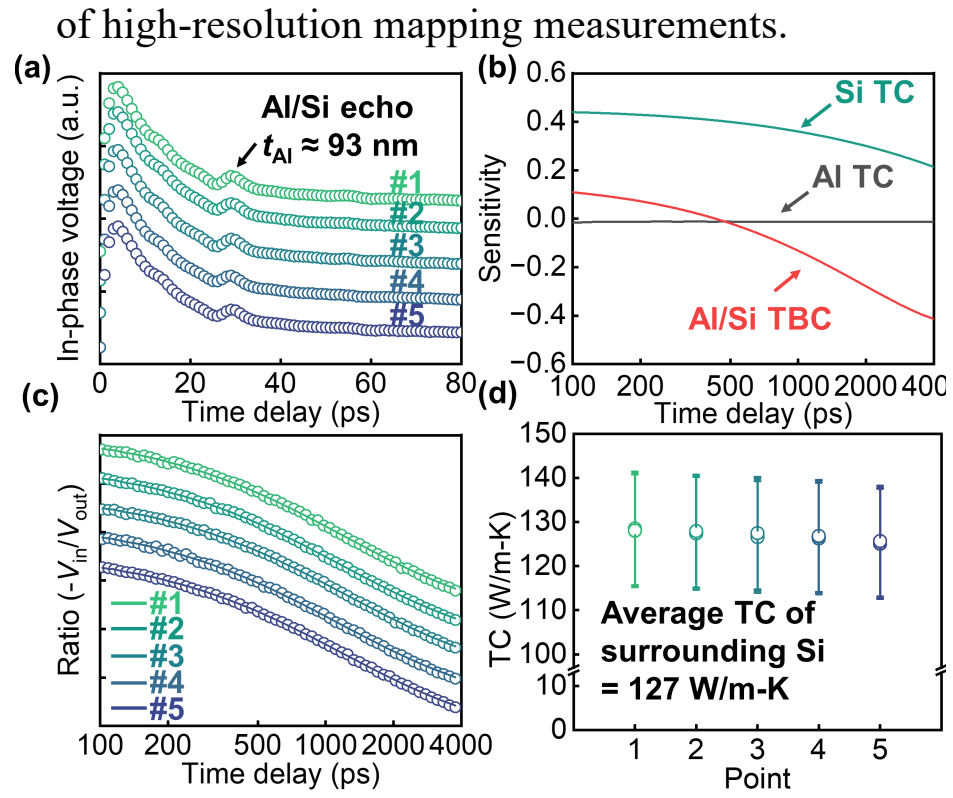

Fig. 7. TDTR measurements of 5 random points in the Si substrate that surrounds TSVs. Overlapping curves are vertically separated for ease of reading. (a) Picosecond ultrasonics measurements of Al transducer thickness. (b) Sensitivities of the TDTR signal to different fitting parameters. (c) TDTR signals (dots) and best fits (lines). (d) Measured TCs of the chosen points.

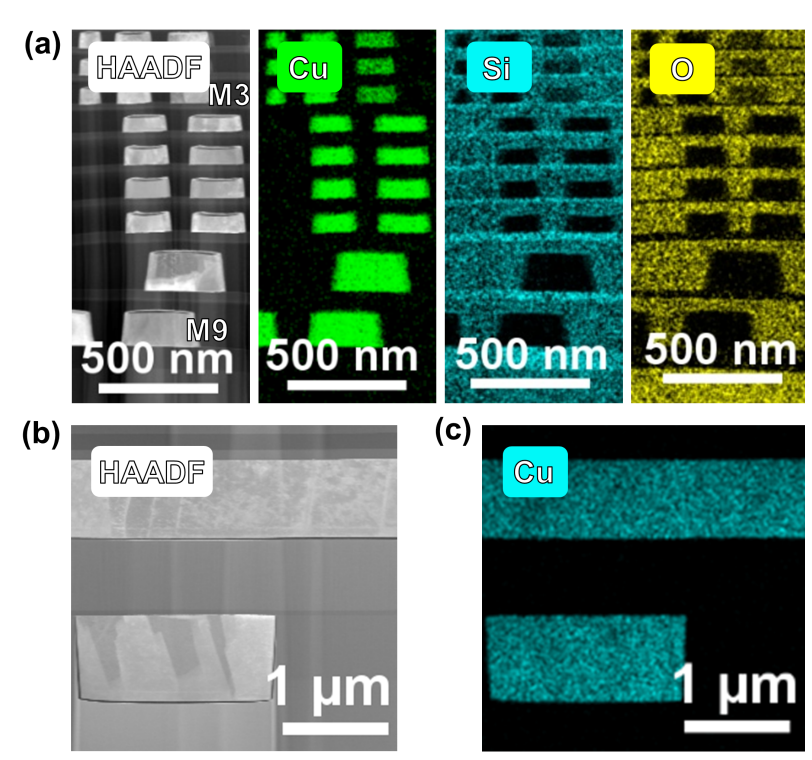

Fig. 8. High angle annular dark-field (HAADF) STEM images and energy-dispersive X-ray spectroscopy (EDS) elemental distribution of (a) thin interconnect layers in Die 1 and (b-c) thick RDL Cu wires in Die 2. These images help identify Cu and ILD regions to extract their thicknesses for subsequent calculation of $C_{V,eff}$.

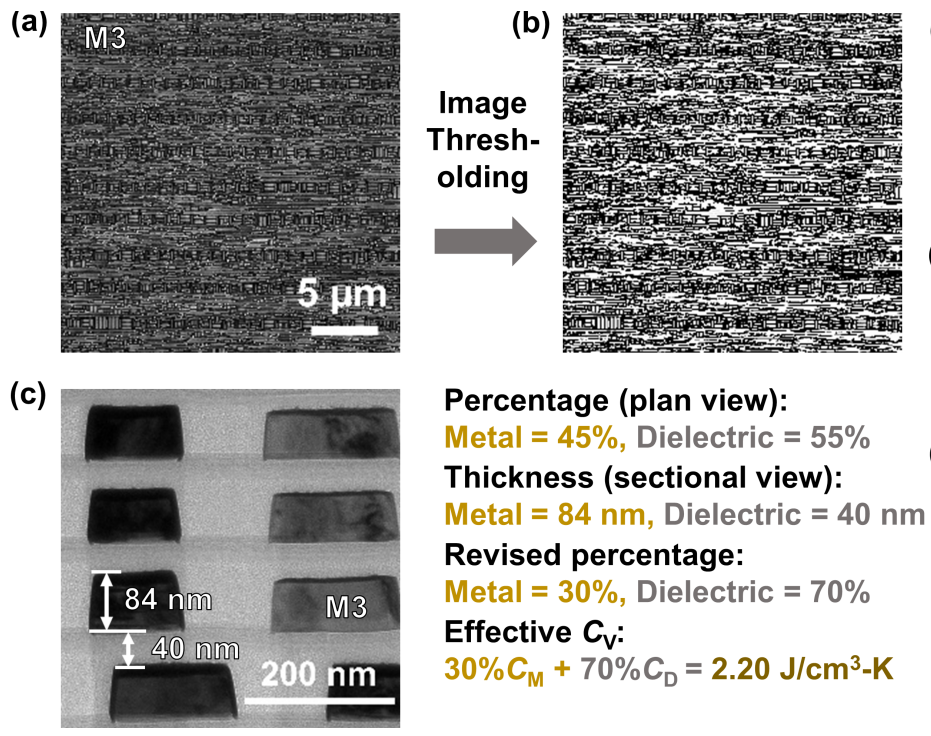

Fig. 9. Example of obtaining the effective volumetric heat capacity parameters ($C_{V,\mathrm{eff}}$) of thin interconnect layers through (a) SEM plan views, (b) thresholded images and (c) the TEM cross-sectional view, which are used in the TDTR data fitting process. The exposed metal layer and the ILD layer below are treated as a mixture of Cu and SiO.

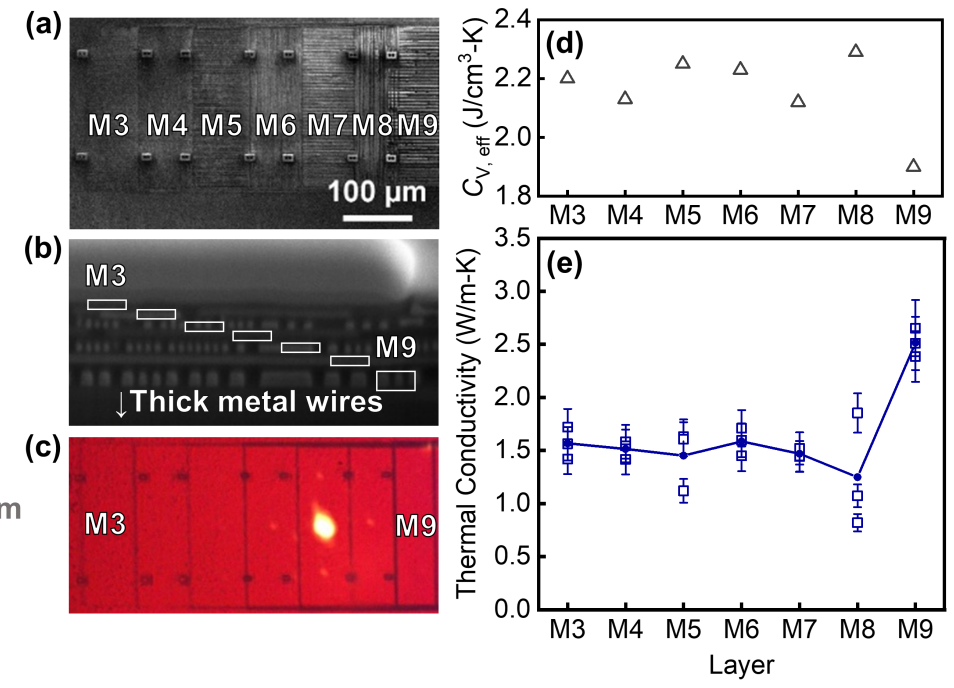

Fig. 10. (a) SEM plan view, (b) FIB/SEM cross-sectional view and (c) CCD image of the interconnect stack prepared via PFIB. The surface of each layer is exposed as a stair-like structure. Note that M3 is the metal layer closest to the active layer. (d) Estimated effective volumetric heat capacity and (e) TDTR-measured TCs of different interconnect layers. Solid marks and line indicate average values.

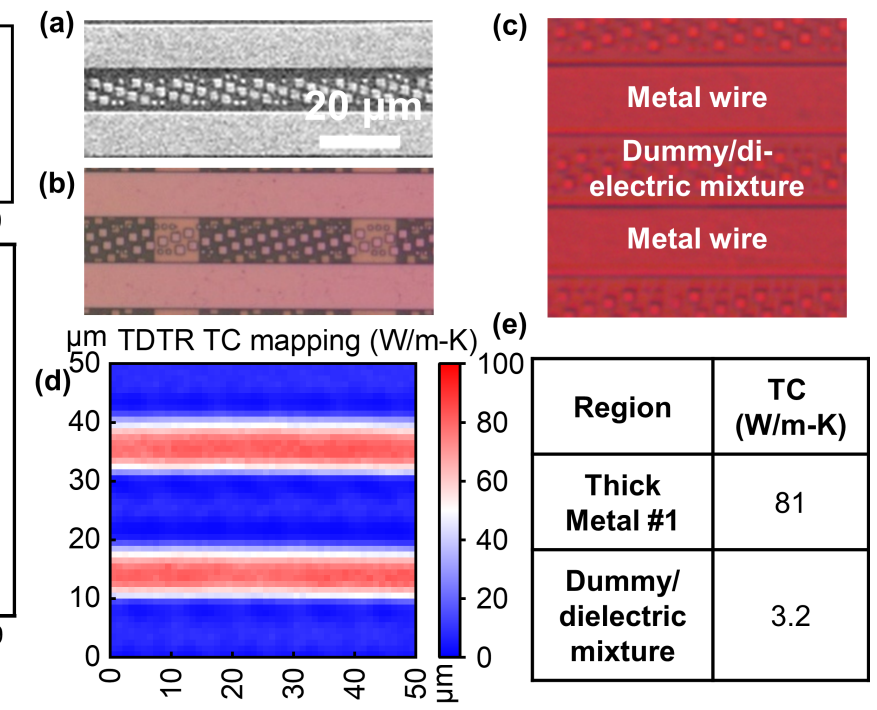

Fig. 11. Thermal characterization of the topmost thick metal layer (numbered #1) in Die 2. (a) SEM and (b) OM images of wires with dummy regions in between. (c) CCD image and (d) visualization of TC distribution for wires and dummy regions by TDTR mapping. (e) Fitted TCs of thick metal and dummy/dielectric mixture from single-point TDTR measurements.

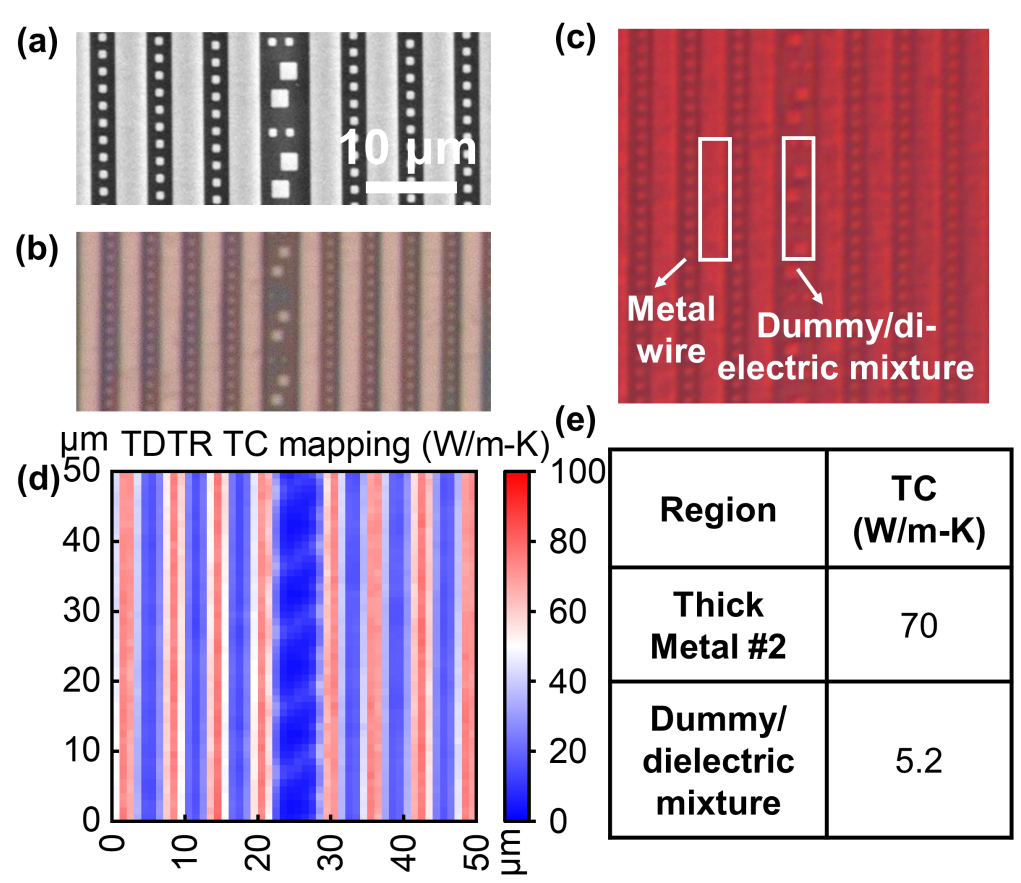

Fig. 12. Thermal characterization of the second highest thick metal layer (numbered #2) in Die 2. (a) SEM and (b) OM images of wires with dummy regions in between. (c) CCD image and (d) visualization of TC mapping for wires and dummy regions by TDTR mapping. (e) Fitted TCs of thick metal and dummy/dielectric mixture from single-point TDTR measurements.

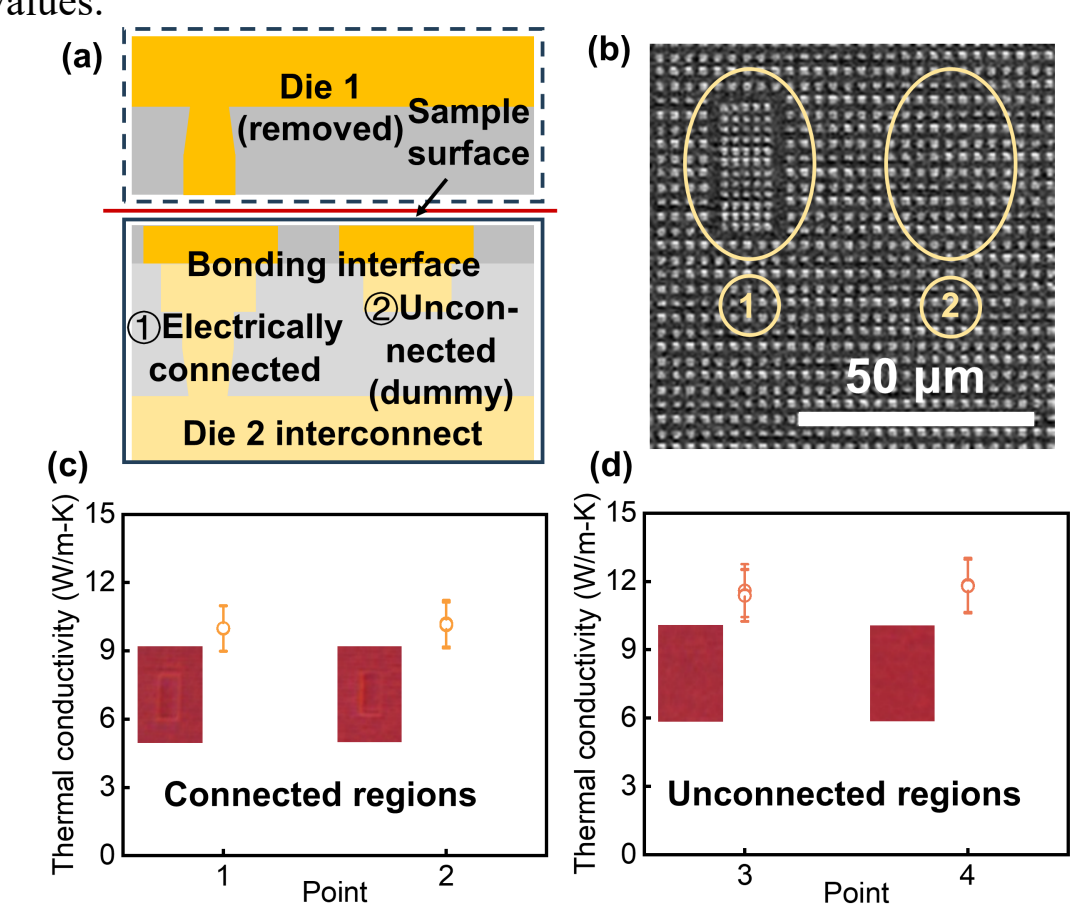

Fig. 13. Thermal characterization of the bonding interface. (a) Schematic and (b) SEM plan-view image of the sample which exposes the lower side of the interface packed with bonding pads, highlighting regions that are electrically connected or isolated. (c-d) TDTR-measured TCs for connected and unconnected regions, each with 2 different points and 3 consecutive measurements for each point.

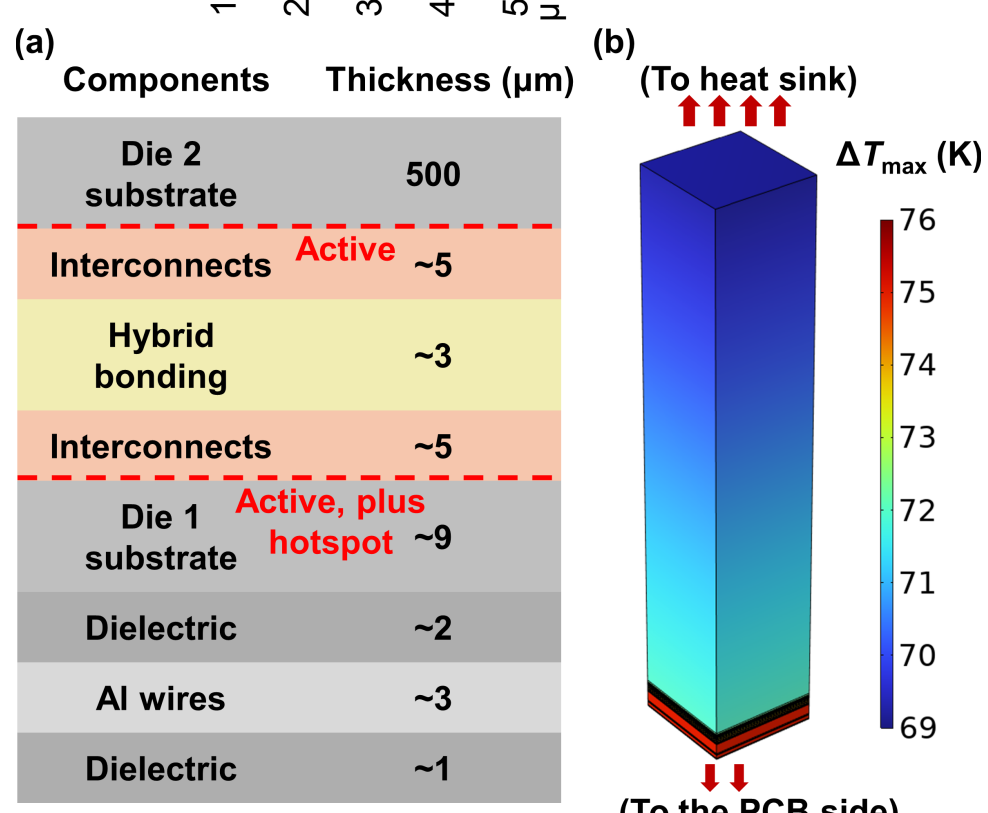

(c)

| Substrate Material | TC (W/m-K) |
|---|---|
| Si | 127 |
| SiC | 400 |
| Diamond | 2000 |

- Size of simulated dies with uniform power densities: 100×100 μm²
- Size of additional hotspot: 10×10 μm²
- Power density of hotspot: 1,000 W/cm²

Fig. 14. Thermal model and simulation based on experimental results. (a) Thicknesses of the stacked layers (not to scale). (b) Temperature rise distribution under a localized hotspot in Die 1 plus uniform power maps of 50 and 10 W/cm² in Dies 1 and 2, respectively. (c) TCs of the substrate materials used in the simulations.

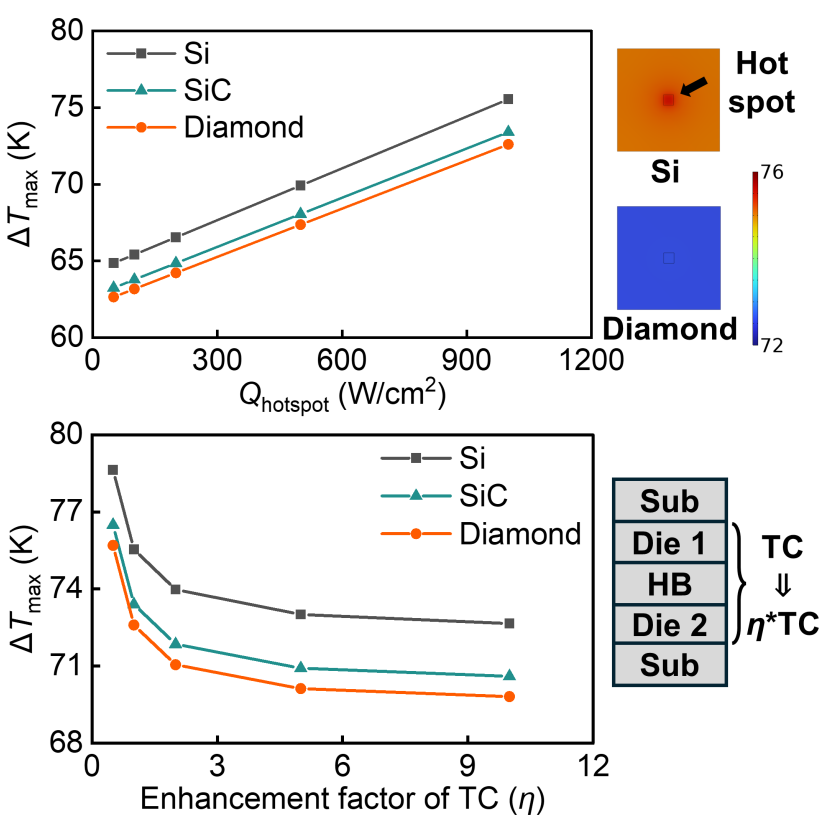

Fig. 15. (a) Calculated effect of different hotspot power densities on the peak temperature rise for Si, SiC, and diamond substrates. The insets are visualizations of the cases with Si and diamond substrates when $Q_{\mathrm{hotspot}}$ is 1,000 W/cm² as in Fig. 14. (b) Calculated effect of TC enhancement of interconnect and bonding layers on the peak temperature rise. The TCs of interconnect and bonding layers are simultaneously multiplied by a factor of $\eta$ to simulate the introduction of high-TC dielectrics.

| Work | Thermal characterization scope | Framework output | Enabled capability |
|---|---|---|---|
| Shen *et al.*[5] | Representative BEOL structures | Thermal properties of BEOL materials | Material thermal analysis |
| Huang *et al.*[3] | Layer-resolved BEOL stacks | 3D thermal conductivity distribution of BEOL | Experiment-based BEOL thermal evaluation |
| He *et al.*[4] | Active layer and interconnects | Local thermal property characterization | Junction thermal analysis |
| Nili *et al.*[12] | Bonded/pre-bonded HB interface | HB thermal resistance/TC mapping and prediction | Process-dependent HB thermal analysis |
| **This work** | **Entire 3D-IC heat conduction pathway** | **Layer-resolved 3D thermal database and thermal characterization framework** | **Experiment-based thermal-aware 3D-IC design and cooling optimization** |

Table I. Comparison of experimental thermal coverage and design capability for advanced IC structures.